\documentclass[10pt, conference]{IEEEtran}

\makeatletter
\def\ps@headings{%
\def\@oddhead{\mbox{}\scriptsize\rightmark \hfil \thepage}%
\def\@evenhead{\scriptsize\thepage \hfil \leftmark\mbox{}}%
\def\@oddfoot{}
\def\@evenfoot{}}
\def\blfootnote{\xdef\@thefnmark{}\@footnotetext}
\makeatother

\usepackage{cite}
\usepackage{url}

\usepackage{graphicx}
\usepackage{algorithm}
\usepackage[noend]{algorithmic}
\usepackage{subfig}
\usepackage{amssymb, amsmath,graphicx,charter, latexsym}
\usepackage{enumerate}

\newtheorem{definition}{Definition}
\newtheorem{lemma}{Lemma}
\newtheorem{corollary}{Corollary}
\newtheorem{theorem}{Theorem}

\def\baselinestretch{0.93}
\begin{document}
\title{Utility Maximization for Delay Constrained QoS in Wireless}
\author{
\IEEEauthorblockN{I-Hong Hou} \IEEEauthorblockA{CSL and Department
of CS\\University of Illinois\\Urbana, IL 61801,
USA\\ihou2@illinois.edu} \and \IEEEauthorblockN{P. R.
Kumar}\IEEEauthorblockA{CSL and Department of ECE\\University of
Illinois\\Urbana, IL 61801, USA\\prkumar@illinois.edu}
 }
\maketitle\blfootnote{This material is based upon work partially
supported by USARO under Contract Nos. W911NF-08-1-0238 and
W-911-NF-0710287, AFOSR under Contract FA9550-09-0121, and NSF under
Contract Nos. CNS-07-21992, ECCS-0701604, CNS-0626584, and
CNS-05-19535. Any opinions, findings, and conclusions or
recommendations expressed in this publication are those of the
authors and do not necessarily reflect the views of the above
agencies. }

\begin{abstract}
This paper studies the problem of utility maximization for clients
with delay based QoS requirements in wireless networks. We adopt a
model used in a previous work that characterizes the QoS
requirements of clients by their delay constraints, channel
reliabilities, and delivery ratio requirements. In this work, we
assume that the \emph{utility} of a client is a function of the
delivery ratio it obtains. We treat the delivery ratio for a client
as a tunable parameter by the access point (AP), instead of a given
value as in the previous work. We then study how the AP should
assign delivery ratios to clients so that the total utility of all
clients is maximized.

We apply the techniques introduced in two previous papers to
decompose the utility maximization problem into two simpler
problems, a $CLIENT$ problem and an $ACCESS$-$POINT$ problem. We
show that this decomposition actually describes a bidding game,
where clients bid for the service time from the AP. We prove that
although all clients behave selfishly in this game, the resulting
equilibrium point of the game maximizes the total utility. In
addition, we also establish an efficient scheduling policy for the
AP to reach the optimal point of the $ACCESS$-$POINT$ problem. We
prove that the policy not only approaches the optimal point but also
achieves some forms of fairness among clients. Finally, simulation
results show that our proposed policy does achieve higher utility
than all other compared policies.
\end{abstract}

\section{Introduction}
\label{section:introduction}

We study how to provide QoS to maximize utility for wireless
clients. We jointly consider the delay constraint and channel
unreliability of each client. The access point (AP) assigns delivery
ratios to clients under the delay and reliability constraints. This
distinguishes our work from most other work on providing QoS where
the delivery ratios to clients are taken as given inputs rather than
tunable parameters.

We consider the scenario where there is one AP that serves a set of
wireless clients. We extend the model proposed in a previous work
\cite{IH09}. This model analytically describes three important
factors for QoS: delay, channel unreliability, and delivery ratio.
The previous work also provides a necessary and sufficient condition
for the demands of the set of clients to be feasible. In this work,
we treat the delivery ratios for clients as variables to be
determined by the AP. We assume that each client receives a certain
amount of \emph{utility} when it is provided a delivery ratio. The
relation between utility and delivery ratio is described by an
\emph{utility function}, which may differ from client to client.
Based on this model, we study the problem of maximizing the total
utility of all clients, under feasibility constraints. We show that
this problem can be formulated as a convex optimization problem.

Instead of solving the problem directly, we apply the techniques
introduced by Kelly \cite{FK97} and Kelly, Maulloo, and Tan
\cite{FK98} to decompose the problem of system utility maximization
into two simpler subproblems that describe the behaviors of the
clients and the AP, respectively. We prove that the utility
maximization problem can be solved by jointly solving the two
simpler subproblems. Further, we describe a bidding game for the
reconciliation between the two subproblems. In this game, clients
bid for service time from the AP, and the AP assigns delivery ratios
to clients according to their bids, to optimize its own subproblem,
under feasibility constraints. Based on the AP's behavior, each
client aims to maximize its own net utility, that is, the difference
between the utility it obtains and the bid it pays. We show that,
while all clients behave selfishly in the game, the equilibrium
point of the game solves the two subproblems jointly, and hence
maximizes the total utility of the system.

We then address how to design a scheduling policy for the AP to
solve its subproblem. We propose a very simple priority based
scheduling algorithm for the AP. This policy requires no information
of the underlying channel qualities of the clients and thus needs no
overhead to probe or estimate the channels. We prove that the
long-term average performance of this policy converges to a single
point, which is in fact the solution to the subproblem for the AP.
Further, we also establish that the policy achieves some forms of
fairness.

Our contribution is therefore threefold. First, we formulate the
problem of system utility maximization as a convex optimization
problem. We then show that this problem is amenable to solution by a
bidding game. Finally, we propose a very simple priority based AP
scheduling policy to solve the AP's subproblem, that can be used in
the bidding iteration to reach the optimal point of the system's
utility maximization problem.

Finally, we conduct simulation studies to verify all the theoretical
results. Simulations show that the performance of the proposed
scheduling policy converges quickly to the optimal value of the
subproblem for AP. Also, by jointly applying the scheduling policy
and the bidding game, we can achieve higher total utility than all
other compared policies.

The rest of the paper is organized as follows: Section
\ref{section:related} reviews some existing related work. Section
\ref{section:model} introduces the model for QoS proposed in
\cite{IH09} and also summarizes some related results. In Section
\ref{section:utility}, we formulate the problem of utility
maximization as a convex programming problem. We also show that this
problem can be decomposed into two subproblems. Section
\ref{section:bidding} describes a bidding game that jointly solves
the two subproblems. One phase of the bidding game consists of each
client selfishly maximizing its own net profit, and the other phase
consists of the AP scheduling client transmissions to optimize its
subproblem. Section \ref{section:scheduling} addresses the
scheduling policy to optimize this latter subproblem. Section
\ref{section:simulation} demonstrates some simulation studies.
Finally, Section \ref{section:conclusion} concludes this paper.

\section{Related Work}
\label{section:related}

There has been a lot of research on providing QoS over wireless
channels. Most of the research has focused on admission control and
scheduling policies. Hou, Borkar, and Kumar \cite{IH09} and Hou and
Kumar \cite{IHH09MobiHoc} have proposed analytical models to
characterize QoS requirements, and have also proposed both admission
control and scheduling policies. Ni, Romdhani, and Turletti
\cite{QN04} provides an overview of the IEEE 802.11 mechanisms and
discusses the limitations and challenges in providing QoS in 802.11.
Gao, Cai, and Ngan \cite{DG05}, Niyato and Hossain \cite{DN05}, and
Ahmed \cite{MHA05} have surveyed existing admission control
algorithms in different types of wireless networks. On the other
hand, Fattah and Leung \cite{HF02} and Cao and Li \cite{YC01} have
provided extensive surveys on scheduling policies for providing QoS.

There is also research on utility maximization for both wireline and
wireless networks. Kelly \cite{FK97} and Kelly, Maulloo, and Tan
\cite{FK98} have considered the rate control algorithm to achieve
maximum utility in a wireline network. Lin and Shroff \cite{XL06}
has studied the same problem with multi-path routing. As for
wireless networks, Xiao, Shroff, and Chong \cite{MX03} has proposed
a power-control framework to maximize utility, which is defined as a
function of the signal-to-interference ratio and cannot reflect
channel unreliability. Cao and Li \cite{YC02} has proposed a
bandwidth allocation policy that also considers channel degradation.
Bianchi, Campbell, and Liao \cite{BG98} has studied utility-fair
services in wireless networks. However, all the aforementioned works
assume that the utility is only determined by the allocated
bandwidth. Thus, they do not consider applications that require
delay bounds.

\section{System Model and Feasibility Condition}
\label{section:model}

We adopt the model proposed in a previous work \cite{IH09} to
capture two key QoS requirements, delay constraints and delivery
ratio requirements, and incorporating channel conditions for users.
In this section, we describe the proposed model and summarize
relevant results of \cite{IH09}.

We consider a system with $N$ clients, numbered as
$\{1,2,\dots,N\}$, and one access point (AP). Packets for clients
arrive at the AP and the AP needs to dispatch packets to clients to
meet their respective requirements. We assume that time is slotted,
with slots numbered as $\{0,1,2,\dots\}$. The AP can make exactly
one transmission in each time slot. Thus, the length of a time slot
would include the times needed for transmitting a DATA packet, an
ACK, and possibly other MAC headers. Assume there is one packet
arriving at the AP periodically for each client, with a fixed period
of $\tau$ time slots, at time slots $0, \tau, 2\tau,\dots$. Each
packet that arrives at the beginning of a period $[k\tau,
(k+1)\tau)$ must be delivered within the ensuing period, or else it
expires and is dropped from the system at the end of this period.
Thus, a delay constraint of $\tau$ time slots is enforced on all
successfully delivered packet. Further, unreliable and heterogeneous
wireless channels to these clients are considered. When the AP makes
a transmission for client $n$, the transmission succeeds (by which
is meant the successful deliveries of both the DATA packet and the
ACK) with probability $p_n$. Due to the unreliable channels and
delay constraint, it may not be possible to deliver the arrived
packets of all the clients. Therefore, each client stipulates a
certain delivery ratio $q_n$ that it has to receive, which is
defined as the average proportion of periods in which its packet is
successfully delivered. The previous work also shows how this model
can be used to capture scenarios where both uplink traffic and
downlink traffic exist.

Below we describe the formal definitions of the concepts of
\emph{fulfilling} a set of clients and the \emph{feasibility} of a
set of client requirements.

\begin{definition}  \label{definition:fufill}
A set of clients with the above QoS constraints is said to be
\emph{fulfilled} by a particular scheduling policy $\eta$ of the AP
if the time averaged delivery ratio of each client is at least $q_n$
with probability 1.
\end{definition}

\begin{definition}  \label{definition:feasible}
A set of clients is \emph{feasible} if there exists some scheduling
policy of the AP that fulfills it.
\end{definition}

Whether a certain client is fulfilled can be decided by the average
number of time slots that the AP spends on working for the client
per period:

\begin{lemma}   \label{lemma:workload}
The delivery ratio of client $n$ converges to $q_n$ with probability
one if and only if the \emph{work performed on client $n$}, defined
as the long-term average number of time slots that the AP spends on
working for client $n$ per period, converges to
$w_n(q_n)=\frac{q_n}{p_n}$ with probability one. We therefore call
$w_n(q_n)$ the \emph{workload} of client $n$.
\end{lemma}

Since expired packets are dropped from the system at the end of each
period, there is exactly one packet for each client at the beginning
of each period. Therefore, there may be occasions where the AP has
delivered all packets before the end of a period and is therefore
forced to stay idle for the remaining time slots in the period. Let
$I_S$ be the expected number of such forced idle time slots in a
period when the client set is just $S \subseteq \{ 1,2, \ldots , N
\}$ (i.e., all clients except those in $S$ are removed from
consideration), and the AP only caters to the subset $S$ of clients.
Since each client $n\in S$ requires $w_n$ time slots per period on
average, we can obtain a necessary condition for feasibility:
$\sum_{i\in S} w_i(q_i) + I_{S} \leq \tau$, for all $S\subseteq
\{1,2,\dots, N\}$. It is shown in \cite{IH09} that this necessary
condition is also sufficient:

\begin{theorem} \label{theorem:feasibility}
A set of clients, with delivery ratio requirements $[q_n]$, is
feasible if and only if $\sum_{i\in S} \frac{q_i}{p_i}\leq \tau -
I_{S}$, for all $S\subseteq \{1,2,\dots, N\}$.
\end{theorem}

\section{Utility Maximization and Decomposition}  \label{section:utility}
In the previous section, it is assumed that the delivery ratio
requirements, $[q_n]$, are given and fixed. In this paper, we
address the problem of how to choose $q:=[q_n]$ so that the total
utility of all the clients in the system can be maximized.

We begin by supposing that each client has a certain utility
function, $U_n(q_n)$, which is strictly increasing, strictly
concave, and continuously differentiable function over the range $0<
q_n\leq 1$, with the value at 0 defined as the right limit, possibly
$-\infty$. The problem of choosing $q_n$ to maximize the total
utility, under the feasibility constraint of Theorem
\ref{theorem:feasibility}, can be described by the following convex
optimization problem:
\\
\\
\textbf{SYSTEM:}
\begin{align}
\mbox{Max }&\mbox{$\sum_{i=1}^{N}$} U_i(q_i) \label{SYSTEM:1}\\
\mbox{s.t. } &\mbox{$\sum_{i\in S}$} \frac{q_i}{p_i} \leq
\tau-I_{S},
\forall S\subseteq \{1,2,\dots, N\},\label{SYSTEM:2}\\
\mbox{over } & q_n\geq 0, \forall 1\leq n\leq N.\label{SYSTEM:3}
\end{align}

It may be difficult to solve $SYSTEM$ directly. So, we decompose it
into two simpler problems, namely, $CLIENT$ and $ACCESS$-$POINT$, as
described below. This decomposition was first introduced by Kelly
\cite{FK97}, though in the context of dealing with rate control for
non-real time traffic.

Suppose client $n$ is willing to pay an amount of $\rho_n$ per
period, and receives a long-term average delivery ratio $q_n$
proportional to $\rho_n$, with $\rho_n = \psi_nq_n$. If $\psi_n>0$,
the utility maximization problem for client $n$ is:
\\
\\
\textbf{CLIENT$_n$:}
\begin{align}
\mbox{Max }&U_n(\frac{\rho_n}{\psi_n})-\rho_n\label{CLIENT:1}\\
\mbox{over } & 0\leq \rho_n\leq \psi_n.\label{CLIENT:2}
\end{align}

On the other hand, given that client $n$ is willing to pay $\rho_n$
per period, we suppose that the AP wishes to find the vector $q$ to
maximize $\sum_{i=1}^N \rho_i\log q_i$, under the feasibility
constraints. In other words, the AP has to solve the following
optimization problem:
\\
\\
\textbf{ACCESS-POINT:}
\begin{align}
\mbox{Max }&\mbox{$\sum_{i=1}^N$} \rho_i\log q_i \label{NETWORK:1}\\
\mbox{s.t. } &\mbox{$\sum_{i\in S}$} \frac{q_i}{p_i} \leq
\tau-I_{S},
\forall S\subseteq \{1,2,\dots, N\},\label{NETWORK:2}\\
\mbox{over } & q_n\geq 0, \forall 1\leq n\leq N.\label{NETWORK:3}
\end{align}

We begin by showing that solving $ACCESS$-$POINT$ is equivalent to
jointly solving $CLIENT_n$ and $ACCESS$-$POINT$.

\begin{theorem} \label{theorem:3to1}
There exist non-negative vectors $q$, $\rho:=[\rho_n]$, and
$\psi:=[\psi_n]$, with $\rho_n=\psi_nq_n$, such that:
\begin{enumerate}[(i)]
\item For $n$ such that $\psi_n>0$, $\rho_n$ is a solution to $CLIENT_n$;\label{3to1:1}
\item Given that each client $n$ pays $\rho_n$ per period, $q$ is a solution to $ACCESS$-$POINT$.  \label{3to1:2}
\end{enumerate}
Further, if $q$, $\rho$, and $\psi$ are all positive vectors, the
vector $q$ is also a solution to $SYSTEM$.
\end{theorem}
\begin{IEEEproof}
We will first show the existence of $q$, $\rho$, and $\psi$ that
satisfy (\ref{3to1:1}) and (\ref{3to1:2}). We will then show that
the resulting $q$ is also the solution to $SYSTEM$.

There exists some $\epsilon>0$ so that by letting $q_n\equiv
\epsilon$, the vector $q$ is an interior point of the feasible
region for both $SYSTEM$ (\ref{SYSTEM:2}) (\ref{SYSTEM:3}), and
$ACCESS$-$POINT$ (\ref{NETWORK:2}) (\ref{NETWORK:3}). Also, by setting
$\rho_n\equiv \epsilon$, $\rho_n$ is also an interior point of the
feasible region for $CLIENT_n$ (\ref{CLIENT:2}). Therefore, by
Slater's condition, a feasible point for $SYSTEM$, $CLIENT_n$, or
$ACCESS$-$POINT$, is the optimal solution for the respective problem
if and only if it satisfies the corresponding Karush-Kuhn-Tucker
(KKT) condition for the problem. Further, since the feasible region
for each of the problems is compact, and the utilities are
continuous on it, or since the utility converges to $-\infty$ at
$q_n=0$, there exists an optimal solution to each of them.

The Lagrangian of $SYSTEM$ is:
\begin{align*}
\begin{array}{rl}
&L_{SYS}(q,\lambda,\nu):=-\sum_{i=1}^{N}
U_i(q_i)\\&+\sum_{S\subseteq \{1,2,\dots, N\}}\lambda_S[\sum_{i\in
S} \frac{q_i}{p_i}-(\tau-I_{S})]-\sum_{i=1}^N \nu_iq_i,
\end{array}
\end{align*}
where $\lambda:=[\lambda_S:S\subseteq \{1,2,\dots, N\}]$ and
$\nu:=[\nu_n:1\leq n\leq N]$ are the Lagrange multipliers. By the
KKT condition, a vector $q^* := [q_1^*, q_2^*,\dots, q_N^*]$ is the
optimal solution to SYSTEM if $q^*$ is feasible and there exists
vectors $\lambda^*$ and $\nu^*$ such that:
\begin{align}
&\begin{array}{ll}\frac{\partial L_{SYS}}{\partial q_n}
\bigg|_{q^*,\lambda^*,\nu^*}&=-U_n'(q_n^*)+\frac{\sum_{S\ni n}
\lambda_S^*}{p_n}-\nu_n^*\\&=0,\forall 1\leq n\leq N,\end{array}\label{SYSTEM:KKT1}\\
&\lambda_S^*[\mbox{$\sum_{i\in S}$}
\frac{q^*_i}{p_i}-(\tau-I_{S})]=0,\forall
S\subseteq \{1,2,\dots, N\},\label{SYSTEM:KKT2}\\
&\nu_n^*q^*_n = 0,\forall 1\leq n\leq N,\label{SYSTEM:KKT3}\\
&\lambda_S^*\geq 0, \forall S\subseteq \{1,\dots, N\}, \mbox{ and
}\nu_n^*\geq 0,\forall 1\leq n\leq N.\label{SYSTEM:KKT4}
\end{align}

The Lagrangian of $CLIENT_n$ is:
\begin{equation*}
L_{CLI}(\rho_n,\xi_n):=-U_n(\frac{\rho_n}{\psi_n})+\rho_n-\xi_n\rho_n,
\end{equation*}
where $\xi_n$ is the Lagrange multiplier for $CLIENT_n$. By the KKT
condition, $\rho_n^*$ is the optimal solution to $CLIENT_n$ if
$\rho_n^*\geq 0$ and there exists $\xi^*_n$ such that:
\begin{align}
&\frac{dL_{CLI}}{d\rho_n}\bigg|_{\rho_n^*,\xi_n^*}=-\frac{1}{\psi_n}U_n'(\frac{\rho_n^*}{\psi_n})+1-\xi_n^*=0,
\label{CLIENT:KKT1}\\
&\xi_n^*\rho_n^*=0,\label{CLIENT:KKT2}\\
&\xi_n^*\geq 0.\label{CLIENT:KKT3}
\end{align}

Finally, the Lagrangian of $ACCESS$-$POINT$ is:
\begin{align*}
\begin{array}{ll}
&L_{NET}(q,\zeta,\mu):=-\sum_{i=1}^{N} \rho_i\log
q_i\\&+\sum_{S\subseteq \{1,2,\dots, N\}}\zeta_S[\sum_{i\in S}
\frac{q_i}{p_i}-(\tau-I_{S})]-\sum_{i=1}^N \mu_iq_i,
\end{array}
\end{align*}
where $\zeta:=[\zeta_S:S\subseteq \{1,2,\dots, N\}]$ and
$\mu:=[\mu_n:1\leq n\leq N]$ are the Lagrange multipliers. Again, by
the KKT condition, a vector $q^* := [q_n^*]$ is the optimal solution
to $ACCESS$-$POINT$ if $q^*$ is feasible and there exists vectors
$\zeta^*$ and $\mu^*$ such that:
\begin{align}
&\begin{array}{ll}\frac{\partial L_{NET}}{\partial q_n}
\bigg|_{q^*,\zeta^*,\mu^*}&=-\frac{\rho_n}{q_n^*}+\frac{\sum_{S\ni
n}
\zeta_S^*}{p_n}-\mu_n^*\\&=0,\forall 1\leq n\leq N,\end{array}\label{NETWORK:KKT1}\\
&\zeta_S^*[\sum_{i\in S} \frac{q^*_i}{p_i}-(\tau-I_{S})]=0,\forall
S\subseteq \{1,2,\dots, N\},\label{NETWORK:KKT2}\\
&\mu_n^*q^*_n = 0,\forall 1\leq n\leq N,\label{NETWORK:KKT3}\\
&\zeta_S^*\geq 0, \forall S\subseteq \{1,\dots, N\}, \mbox{ and
}\mu_n^*\geq 0,\forall 1\leq n\leq N.\label{NETWORK:KKT4}
\end{align}

Let $q^*$ be a solution to $SYSTEM$, and let $\lambda^*$, $\nu^*$ be
the corresponding Lagrange multipliers that satisfy conditions
(\ref{SYSTEM:KKT1})--(\ref{SYSTEM:KKT4}). Let $q_n=q_n^*$,
$\rho_n=\frac{\sum_{S\ni n} \lambda_S^*}{p_n}q_n^*$, and
$\psi_n=\frac{\sum_{S\ni n} \lambda_S^*}{p_n}$, for all $n$.
Clearly, $q$, $\rho$, and $\psi$ are all non-negative vectors. We
will show $(q,\rho,\psi)$ satisfy (\ref{3to1:1}) and (\ref{3to1:2}).

We first show (\ref{3to1:1}) for all $n$ such that
$\psi_n=\frac{\sum_{S\ni n} \lambda_S^*}{p_n}>0$. It is obvious that
$\rho_n=\psi_nq_n$. Also, $\rho_n\geq0$, since $\lambda_S^*\geq 0$
(by (\ref{SYSTEM:KKT4})) and $q_n^*\geq 0$ (since $q^*$ is
feasible). Further, let the Lagrange multiplier of $CLIENT_n$,
$\xi_n$, be equal to $\nu_n^*/\frac{\sum_{S\ni n}
\lambda_S^*}{p_n}=\nu_n^*/\psi_n$. Then we have:
\begin{align*}
&\begin{array}{ll}
\frac{\partial L_{CLI}}{\partial \rho_n}&\bigg|_{\rho_n,\xi_n}=-\frac{1}{\psi_n}U_n'(\frac{\rho_n}{\psi_n})+1-\xi_n\\
&=\frac{1}{\psi_n}(-U_n'(\frac{\rho_n}{\psi_n})+\psi_n-\psi_n\xi_n)\\
&=\frac{1}{\psi_n}(-U_n'(q_n^*)+\frac{\sum_{S\ni
n}\lambda_S^*}{p_n}-\nu_n^*) = 0,\mbox{ by (\ref{SYSTEM:KKT1}),}
\end{array}\\
&\xi_n\rho_n=\frac{\nu_n^*}{\psi_n}\psi_nq_n^* = \nu_n^*q_n^*=0,\mbox{ by (\ref{SYSTEM:KKT3})} \\
&\xi_n=\nu_n^*/\frac{\sum_{S\ni n} \lambda_S^*}{p_n}\geq 0,\mbox{ by
(\ref{SYSTEM:KKT4})}.
\end{align*}
In sum, $(\rho,\psi,\xi)$ satisfies the KKT conditions for
$CLIENT_n$, and therefore $\rho_n$ is a solution to $CLIENT_n$, with
$\rho_n=\psi_nq_n$.

Next we establish (\ref{3to1:2}). Since $q=q^*$ is the solution to
$SYSTEM$, it is feasible. Let the Lagrange multipliers of
$ACCESS$-$POINT$ be $\zeta_S = \lambda_S^*,\forall S$, and
$\mu_n=0,\forall n$, respectively. Given that each client $n$ pays
$\rho_n$ per period, we have:
\begin{align*}
&\begin{array}{ll}\frac{\partial L_{NET}}{\partial q_n}
\bigg|_{q,\zeta,\mu}&=-\frac{\rho_n}{q_n}+\frac{\sum_{S\ni n}
\zeta_S}{p_n}-\mu_n\\&=-\psi_n+\psi_n-0=0,\forall n,\end{array}\\
&\begin{array}{ll}\zeta_S[\sum_{i\in S}
\frac{q_i}{p_i}-(\tau-I_{S})]&=\lambda_S^*[\sum_{i\in S}
\frac{q_i^*}{p_i}-(\tau-I_{S})]\\&=0,\forall
S,\mbox{ by (\ref{SYSTEM:KKT2})},\end{array}\\
&\mu_nq_n = 0\times q_n = 0,\forall n,\\
&\zeta_S=\lambda_S^*\geq 0, \forall S\mbox{ (by (\ref{SYSTEM:KKT4})),
and }\mu_n\geq 0,\forall n.
\end{align*}
Therefore, $(q,\zeta,\mu)$ satisfies the KKT condition for $ACCESS$-$POINT$
and thus $q$ is a solution to $ACCESS$-$POINT$.

For the converse, suppose $(q,\rho,\psi)$ are positive vectors with
$\rho_n=\psi_nq_n$, for all $n$, that satisfy (\ref{3to1:1}) and
(\ref{3to1:2}). We wish to show that $q$ is a solution to $SYSTEM$.
Let $\xi_n$ be the Lagrange multiplier for $CLIENT_n$. Since we
assume $\psi_n>0$ for all $n$, the problem $CLIENT_n$ is
well-defined for all $n$, and so is $\xi_n$. Also, let $\zeta$ and
$\mu$ be the Lagrange multipliers for $ACCESS$-$POINT$. Since $q_n>0$
for all $n$, we have $\mu_n=0$ for all $n$ by (\ref{NETWORK:KKT3}).
By (\ref{NETWORK:KKT1}), we also have:
\begin{align*}
\begin{array}{ll}
\frac{\partial L_{NET}}{\partial q_n}
\bigg|_{q,\zeta,\mu}&=-\frac{\rho_n}{q_n}+\frac{\sum_{S\ni n}
\zeta_S}{p_n}-\mu_n\\
&=-\psi_n +\frac{\sum_{S\ni n} \zeta_S}{p_n} =0,
\end{array}
\end{align*}
and thus $\psi_n=\frac{\sum_{S\ni n} \zeta_S}{p_n}$. Let
$\lambda_S=\zeta_S$, for all $S$, and $\nu_n=\psi_n\xi_n$, for all
$n$. We claim that $q$ is the optimal solution to $SYSTEM$ with
Lagrange multipliers $\lambda$ and $\nu$.

Since $q$ is a solution to $ACCESS$-$POINT$, it is feasible. Further, we
have:
\begin{align*}
&\begin{array}{ll}\frac{\partial L_{SYS}}{\partial q_n}
\bigg|_{q,\lambda,\nu}&=-U_n'(q_n)+\frac{\sum_{S\ni n}
\lambda_S}{p_n}-\nu_n\\&=-U_n'(\frac{\rho_n}{\psi_n})+\psi_n-\psi_n\xi_n=0,\forall n, \mbox{ by (\ref{CLIENT:KKT1})},\end{array}\\
&\begin{array}{ll}\lambda_S[\sum_{n\in S}
\frac{q_n}{p_n}-(\tau-I_{S})]&=\zeta_S[\sum_{n\in S}
\frac{q_n}{p_n}-(\tau-I_{S})]\\&=0,\forall
S,\mbox{ by (\ref{NETWORK:KKT2})},\end{array}\\
&\nu_nq_n = \xi_n\rho_n=0,\forall n,\mbox{ by (\ref{CLIENT:KKT2})},\\
&\lambda_S=\zeta_S\geq 0, \forall S,\mbox{ by (\ref{NETWORK:KKT4}),}\\
&\nu_n=\psi_n\xi_n\geq 0,\forall n,\mbox{ by
(\ref{CLIENT:KKT3})}.
\end{align*}
Thus, $(q,\lambda,\nu)$ satisfy the KKT condition for $SYSTEM$, and
so $q$ is a solution to $SYSTEM$.
\end{IEEEproof}

\section{A Bidding Game between Clients and Access Point}    \label{section:bidding}

Theorem \ref{theorem:3to1} states that the maximum total utility of
the system can be achieved when the solutions to the problems
$CLIENT_n$ and $ACCESS$-$POINT$ agree. In this section, we formulate
a repeated game for such reconciliation. We also discuss the
meanings of the problems $CLIENT_n$ and $ACCESS$-$POINT$ in this
repeated game.

The repeated game is formulated as follows:

\begin{enumerate}[Step 1:]
\item Each client $n$ announces an amount $\rho_n$ that it pays per period.
\item After noting the amounts, $\rho_1,\rho_2,\dots, \rho_N$, paid by the clients, the AP
chooses a scheduling policy so that the resulting long-term delivery
ratio, $q_n$, for each client maximizes $\sum_{i=1}^N \rho_i\log
q_i$.
\item The client $n$ observes its own delivery ratio,
$q_n$. It computes $\psi_n:=\rho_n/q_n$. It then determines
$\rho_n^*\geq 0$ to maximize
$U_n(\frac{\rho_n^*}{\psi_n})-\rho_n^*$. Client $n$ updates the
amount it pays to $(1-\alpha)\rho_n+\alpha \rho_n^*$, with some
fixed $0<\alpha<1$, and announces the new bid value.
\item Go back to Step 2.
\end{enumerate}

In Step 3 of the game, client $n$ chooses its new amount of payment
as a weighted average of the past amount and the derived optimal
value, instead of the derived optimal value. This design serves two
purposes. First, it seeks to avoid the system from oscillating
between two extreme values. Second, since $\rho_n$ is initiated to a
positive value, and $\rho_n^*$ derived in each iteration is always
non-negative, this design guarantees $\rho_n$ to be positive
throughout all iterations. Since $\psi_n=\rho_n/q_n$, this also
ensures $\psi_n>0$ and the function $U_n(\frac{\rho_n}{\psi_n})$ is
consequently always well-defined.

We show that the fixed point of this repeated game maximizes the
total utility of the system:

\begin{theorem} \label{theorem:converge}
Suppose at the fixed point of the repeated game, each client $n$
pays $\rho_n^*$ per period, and receives delivery ratio $q_n^*$. If
both $\rho_n^*$ and $q_n^*$ are positive for all $n$, the vector
$q^*$ maximizes the total utility of the system.
\end{theorem}
\begin{IEEEproof}
Let $\psi_n^*=\frac{\rho_n^*}{q_n^*}$. It is positive since both
$\rho_n^*$ and $q_n^*$ are positive. Since the vectors $q^*$ and
$\rho^*$ are derived from the fixed point, $\rho_n^*$ maximizes
$U_n(\frac{\rho_n}{\psi_n^*})-\rho_n$, over all $\rho_n\geq 0$, as
described in Step 3 of the game. Thus, $\rho_n^*$ is a solution to
$CLIENT_n$, given $\rho_n^*=\psi_n^*q_n^*$. Similarly, from Step 2,
$q^*$ is the feasible vector that maximizes $\sum_{i=1}^N
\rho_i^*\log q_i$, over all feasible vectors $q$. Thus, $q^*$ is a
solution to $ACCESS$-$POINT$, given that each client $n$ pays
$\rho_n^*$ per period. By Theorem \ref{theorem:3to1}, $q^*$ is the
unique solution to $SYSTEM$ and therefore maximizes the total
utility of the system.
\end{IEEEproof}

Next, we describe the meaning of the game. In Step 3, client $n$
assumes a linear relation between the amount it pays, $\rho_n$, and
the delivery ratio it receives, $q_n$. To be more exactly, it
assumes $\rho_n=\psi_nq_n$, where $\psi_n$ is the price. Thus,
maximizing $U_n(\frac{\rho_n}{\psi_n})-\rho_n$ is equivalent to
maximizing $U_n(q_n)-\rho_n$. Recall that $U_n(q_n)$ is the utility
that client $n$ obtains when it receives delivery ratio $q_n$.
$U_n(q_n)-\rho_n$ is therefore the net profit that client $n$ gets.
In short, in Step 3, the goal of client $n$ is to selfishly maximize
its own net profit using a first order linear approximation to the
relation between payment and delivery ratio.

We next discuss the behavior of the AP in Step 2. The AP schedules
clients so that the resulting delivery ratio vector $q$ is a
solution to the problem $ACCESS$-$POINT$, given that each client $n$
pays $\rho_n$ per period. Thus, $q$ is feasible and there exists
vectors $\zeta$ and $\mu$ that satisfy conditions
(\ref{NETWORK:KKT1})--(\ref{NETWORK:KKT4}). While it is difficult to
solve this problem, we consider a special restrictive case that
gives us a simple solution and insights into the AP's behavior. Let
$TOT:=\{1,2,\dots, N\}$ be the set that consists of all clients. We
assume that a solution $(q,\zeta,\mu)$ to the problem has the
following properties: $\zeta_S=0$, for all $S\neq TOT$,
$\zeta_{TOT}>0$, and $\mu_n=0$, for all $n$. By
(\ref{NETWORK:KKT1}), we have:
\begin{align*}
-\frac{\rho_n}{q_n}+\frac{\sum_{S\ni n}
\zeta_S}{p_n}-\mu_n=-\frac{\rho_n}{q_n}+\frac{\zeta_{TOT}}{p_n}=0,
\end{align*}
and therefore $q_n=p_n\rho_n/\zeta_{TOT}$. Further, since
$\zeta_{TOT}>0$, (\ref{NETWORK:KKT2}) requires that:
\begin{align*}
\sum_{i\in TOT}\frac{q_i}{p_i} - (\tau-I_{TOT})=\sum_{i\in
TOT}\frac{\rho_i}{\zeta_{TOT}} - (\tau-I_{TOT})=0.
\end{align*}
Thus, $\zeta_{TOT}=\frac{\sum_{i=1}^N\rho_i}{\tau-I_{TOT}}$ and
$\frac{q_n}{p_n}=\frac{\rho_n}{\sum_{i=1}^N m_i}(\tau-I_{TOT})$, for
all $n$. Notice that the derived $(q,\zeta,\mu)$ satisfies
conditions (\ref{NETWORK:KKT1})--(\ref{NETWORK:KKT4}). Thus, under
the assumption that $q$ is feasible, this special case actually
maximizes $\sum_{i=1}^N \rho_i\log q_i$. In Section
\ref{section:scheduling} we will address the general situation
without any such assumption, since it needs not be true.

Recall that $I_{TOT}$ is the average number of time slots that the
AP is forced to be idle in a period after it has completed all
clients. Also, by Lemma \ref{lemma:workload}, $\frac{q_n}{p_n}$ is
the workload of client $n$, that is, the average number of time
slots that the AP should spend working for client $n$. Thus, letting
$\frac{q_n}{p_n}=\frac{\rho_n}{\sum_{i=1}^N \rho_i}(\tau-I_{TOT})$,
for all $n$, the AP tries to allocate those non-idle time slots so
that the average number of time slots each client gets is
proportional to its payment. Although we only study the special case
of $I_{TOT}$ here, we will show that the same behavior also holds
for the general case in the Section \ref{section:scheduling}.

In summary, the game proposed in this section actually describes a
bidding game, where clients are bidding for non-idle time slots.
Each client gets a share of time slots that is proportional to its
bid. The AP thus assigns delivery ratios, based on which the clients
calculate a price and selfishly maximize their own net profits.
Finally, Theorem \ref{theorem:converge} states that the equilibrium
point of this game maximizes the total utility of the system.

\section{A Scheduling Policy for Solving $ACCESS$-$POINT$}    \label{section:scheduling}
In Section \ref{section:bidding}, we have shown that by setting
$q_n=p_n\frac{\rho_n}{\sum_{i=1}^N m_i}(\tau-I_{TOT})$, the
resulting vector $q$ solves $ACCESS$-$POINT$ provided $q$ is indeed
feasible. Unfortunately, such $q$ is not always feasible and solving
$ACCESS$-$POINT$ is in general difficult. Even for the special case
discussed in Section \ref{section:bidding}, solving $ACCESS$-$POINT$
requires knowledge of channel conditions, that is, $p_n$. In this
section, we propose a very simple priority based scheduling policy
that can achieve the optimal solution for $ACCESS$-$POINT$, and that
too without any knowledge of the channel conditions.

In the special case discussed in Section \ref{section:bidding}, the
AP tries, though it may be impossible in general, to allocate
non-idle time slots to clients in proportion to their payments.
Based on this intuitive guideline, we design the following
scheduling policy. Let $u_n(t)$ be the number of time slots that the
AP has allocated for client $n$ up to time $t$. At the beginning of
each period, the AP sorts all clients in increasing order of
$\frac{u_n(t)}{\rho_n}$, so that
$\frac{u_1(t)}{\rho_1}\leq\frac{u_2(t)}{\rho_2}\leq\dots$ after
renumbering clients if necessary. The AP then schedules
transmissions according to the priority ordering, where clients with
smaller $\frac{u_n(t)}{\rho_n}$ get higher priorities. Specifically,
in each time slot during the period, the AP chooses the smallest $i$
for which the packet for client $i$ is not yet delivered, and then
transmits the packet for client $i$ in that time slot. We call this
the \emph{weighted transmission} policy (WT). Notice that the policy
only requires the AP to keep track of the bids of clients and the
number of time slots each client has been allocated in the past,
followed by a sorting of $\frac{u_n(t)}{\rho_n}$ among all clients.
Thus, the policy requires no information on the actual channel
conditions, and is tractable. Simple as it is, we show that the
policy actually achieves the optimal solution for $ACCESS$-$POINT$.
In the following sections, we first prove that the vector of
delivery ratios resulting from the WT policy converges to a single
point. We then prove that this limit is the optimal solution for
$ACCESS$-$POINT$. Finally, we establish that the WT policy
additionally achieves some forms of fairness.

\subsection{Convergence of the Weighted Transmission Policy}  \label{subsection:convergence}

We now prove that, by applying the WT policy, the delivery ratios of
clients will converge to a vector $q$. To do so, we actually prove
the convergence property and precise limit of a more general class
of scheduling policies, which not only consists of the WT policy but
also a scheduling policy proposed in \cite{IH09}. The proof is
similar to that used in \cite{IH09} and is based on Blackwell's
approachability theorem \cite{DB56}. The proof in \cite{IH09} only
shows that the vector of delivery ratios approaches a desirable set
in the $N$-space under a particular policy, while here we prove that
the vector of delivery ratios converges to a single point under a
more general class of scheduling policies. Thus, our result is both
stronger and more general than the one in \cite{IH09}.

We start by introducing Blackwell's approachability theorem.
Consider a single player repeated game. In each round $i$ of the
game, the player chooses some action, $a(i)$, and receives a reward
$v(i)$, which is a random vector whose distribution is a function of
$a(i)$. Blackwell studies the long-term average of the rewards
received, $\lim_{j\rightarrow\infty}\sum_{i=1}^j v(i)/j$, defining a
set as \emph{approachable}, under some policy, if the distance
between $\sum_{i=1}^j v(i)/j$ and the set converges to 0 with
probability one, as $j\rightarrow\infty$.

\begin{theorem}[Blackwell \cite{DB56}] \label{theorem:blackwell}
Let $A\subseteq \mathbb{R}^N$ be any closed set. Suppose that for
every $x\notin A$, a policy $\eta$ chooses an action $a$ $(=a(x))$,
which results in an expected payoff vector $E(v)$. If the hyperplane
through $y$, the closest point in $A$ to $x$, perpendicular to the
line segment $xy$, separates $x$ from $E(v)$, then $A$ is
approachable with the policy $\eta$.
\end{theorem}

Now we formulate our more general class of scheduling policies. We
call a policy a \emph{generalized transmission time policy} if, for
a choice of a positive parameter vector $a$ and non-negative
parameter vector $b$, the AP sorts clients by $a_nu_n(t)-b_nt$ at
the beginning of each period, and gives priorities to clients with
lower values of this quantity. Note that the special case $a_n
\equiv \frac{1}{\rho_n}$ and $b_n\equiv0$ yields the WT policy,
while the choice $a_n\equiv 1$ and $b_n\equiv\frac{q_n}{p_n}$ yields
the largest time-based debt first policy of \cite{IH09}, and thus we
describe a more general set of policies.

\begin{theorem} \label{theorem:convergence}
For each generalized transmission time policy, there exists a vector
$q$ such that the vector of work loads resulting from the policy
converges to $w(q):=[w_n(q_n)]$.
\end{theorem}
\begin{IEEEproof}
Given the parameters $\{(a_n,b_n):1\leq n\leq N\}$, we give an exact
expression for the limiting $q$. We define a sequence of sets
$\{H_k\}$ and corresponding values $\{\theta_k\}$ iteratively as
follows. Let $H_0:=\phi$, $\theta_0:=-\infty$, and
\begin{align*}
H_k:=&\arg \min_{S:S\supsetneqq
H_{k-1}}\frac{\frac{1}{\tau}(I_{H_{k-1}}-I_{S})-\sum_{n\in
S\backslash H_{k-1}}\frac{b_n}{a_n}}{\sum_{n\in S\backslash
H_{k-1}}1/a_n},\\
\theta_k:=&\frac{\frac{1}{\tau}(I_{H_{k-1}}-I_{H_k})-\sum_{n\in
{H_k\backslash H_{k-1}}}\frac{b_n}{a_n}}{\sum_{n\in {H_k\backslash
H_{k-1}}}1/a_n}, \mbox{ for all }k>0.
\end{align*}
In selecting $H_k$, we always choose a maximal subset, breaking ties
arbitrary. $(H_1,\theta_1),(H_2,\theta_2),\dots$, can be iteratively
defined until every client is in some $H_k$. Also, by the
definition, we have $\theta_k
> \theta_{k-1}$, for all $k>0$. If client $n$ is in
$H_k\backslash H_{k-1}$, we define $q_n:=\tau
p_n\frac{b_n+\theta_k}{a_n}$, and so
$w_n(q_n)=\tau\frac{b_n+\theta_k}{a_n}$. The proof of convergence
consists of two parts. First we prove that the vector of work
performed (see Lemma \ref{lemma:workload} for definition) approaches
the set $\{w^*|w^*_n\geq w_n(q_n)\}$. Then we prove that $w(q)$ is
the only feasible vector in the set $\{w^*|w^*_n\geq w_n(q_n)\}$.
Since the \emph{feasible region} for work loads, defined as the set
of all feasible vectors for work loads, is approachable under any
policy, the vector of work performed resulting from the generalized
transmission time policy must converge to $w(q)$.

For the first part, we prove the following statement: for each
$k\geq1$, the set $W_k:=\{w^*|w^*_n\geq\tau
\frac{b_n+\theta_k}{a_n}, \forall n\notin H_{k-1}\}$ is
approachable. Since $\cap_{i\geq 0} W_i = \{w^*|w^*_n\geq
w_n(q_n)\}$, we also prove that $\{w^*|w^*_n\geq w_n(q_n)\}$ is
approachable.

Consider a linear transformation on the space of workloads
$L(w):=[l_n:l_n=\frac{a_nw_n/\tau-b_n}{\sqrt{a_n}}]$. Proving $W_k$
is approachable is equivalent to proving that its image under $L$,
$V_k:=\{l|l_n\geq \frac{\theta_k}{\sqrt{a_n}},\forall n\notin
H_{k-1}\}$, is approachable. Now we apply Blackwell's theorem.
Suppose at some time $t$ that is the beginning of a period, the
number of time slots that the AP has worked on client $n$ is
$u_n(t)$. The work performed for client $n$ is
$\frac{u_n(t)}{t/\tau}$, and the image of the vector of work
performed under $L$ is
$x(t):=[x_n(t)|x_n(t)=\frac{a_nu_n(t)/t-b_n}{\sqrt{a_n}}]$, which we
shall suppose is not in $V_k$. The generalized transmission time
policy sorts clients so that $a_1u_1(t)-b_1\leq
a_2u_2(t)-b_2\leq\dots$, or equivalently,
$\sqrt{a_1}x_1(t)\leq\sqrt{a_2}x_2(t)\leq\dots$. The closest point
in $V_k$ to $x(t)$ is $y:=[y_n]$, where
$y_n=\frac{\theta_k}{\sqrt{a_n}}$, if
$x_n(t)<\frac{\theta_k}{\sqrt{a_n}}$ and $n\notin H_{k-1}$, and
$y_n=x_n$, otherwise. The hyperplane that passes through $y$ and is
orthogonal to the line segment $xy$ is:
\begin{align*}
\{z|f(z):=\sum_{n:n\leq n_0,n\notin
H_{k-1}}(z_n-\frac{\theta_k}{\sqrt{a_n}})(x_n(t)-\frac{\theta_k}{\sqrt{a_n}})=0\}.
\end{align*}

Let $\pi_n$ be the expected number of time slots that the AP spends
on working for client $n$ in this period under the generalized
transmission time policy. The image under $L$ of the expected reward
in this period is $\pi_L:=[\frac{a_n\pi_n/\tau-b_n}{\sqrt{a_n}}]$.
Blackwell's theorem shows that $V_k$ is approachable if $x(t)$ and
$\pi_L$ are separated by the plane $\{z|f(z)=0\}$. Since
$f(x(t))\geq 0$, it suffices to show $f(\pi_L)\leq 0$.

We manipulate the original ordering, for this period, so that all
clients in $H_{k-1}$ have higher priorities than those not in
$H_{k-1}$, while preserving the relative ordering between clients
not in $H_{k-1}$. Note this manipulation will not give any client
$n\notin H_{k-1}$ higher priority than it had in the original
ordering. Therefore, $\pi_n$ will not increase for any $n\notin
H_{k-1}$. Since the value of $f(\pi_L)$ only depends on $\pi_n$ for
$n\notin H_{k-1}$, and increases as those $\pi_n$ decrease, this
manipulation will not decrease the value of $f(\pi_L)$. Thus, it
suffices to prove that $f(\pi_L)\leq 0$, under this new ordering.
Let $n_0:=|H_{k-1}|+1$. Under this new ordering, we have:
$\sqrt{a_{n_0}}x_{n_0}(t)\leq
\sqrt{a_{n_0+1}}x_{n_0+1}(t)\leq\dots\leq
\sqrt{a_{n_1}}x_{n_1}(t)<\theta_k\leq
\sqrt{a_{n_1+1}}x_{n_1+1}(t)\leq\dots.$

Let $\delta_n=\sqrt{a_{n}}x_{n}(t)-\sqrt{a_{n+1}}x_{n+1}(t)$, for
$n_0\leq n \leq n_1-1$ and
$\delta_{n_1}=\sqrt{a_{n_1}}x_{n_1}(t)-\theta_k$. Clearly,
$\delta_n\leq 0$, for all $n_0\leq n\leq n_1$. Now we can derive:
\begin{align*}
f(\pi_L)&=\sum_{n=n_0}^{n_1}(\frac{a_n\pi_n/\tau-b_n}{\sqrt{a_n}}-\frac{\theta_k}{\sqrt{a_n}})(x_n(t)-\frac{\theta_k}{\sqrt{a_n}})\\
&=\sum_{n=n_0}^{n_1}(\frac{\pi_n}{\tau}-\frac{b_n}{a_n}-\frac{\theta_k}{a_n})(\sqrt{a_n}x_n(t)-\theta_k)\\
&=\sum_{i=n_0}^{n_1}(\frac{\sum_{n=n_0}^i\pi_n}{\tau}-\sum_{n=n_0}^i\frac{b_n}{a_n}-\theta_k\sum_{n=n_0}^i\frac{1}{a_n})\delta_i.
\end{align*}

Recall that $I_S$ is the expected number of idle time slots when the
AP only caters on the subset $S$. Thus, under this ordering, we have
$\sum_{n=1}^i\pi_n=\tau-I_{\{1,\dots,i\}}$, for all $i$, and
$\sum_{n=n_0}^i\pi_n=I_{\{1,\dots,n_0-1\}}-I_{\{1,\dots,i\}}=I_{H_{k-1}}-I_{\{1,\dots,i\}}$,
for all $i\geq n_0$. By the definition of $H_k$ and $\theta_k$, we
also have
\begin{align*}
&\frac{\sum_{n=n_0}^i\pi_n}{\tau}-\sum_{n=n_0}^i\frac{b_n}{a_n}-\theta_k\sum_{n=n_0}^i\frac{1}{a_n}\\
=&(\sum_{n=n_0}^i\frac{1}{a_n})(\frac{\frac{1}{\tau}(I_{H_{k-1}}-I_{\{1,\dots,i\}})-\sum_{n=n_0}^{i}\frac{b_n}{a_n}}{\sum_{n\in
{\{1,\dots,i\}\backslash H_{k-1}}}1/a_n}-\theta_k)\geq 0.
\end{align*}
Therefore, $f(\pi_L)\leq 0$, since $\delta_i\leq0$, and $V_k$ is
indeed approachable, for all $k$.

We have established that the set $\{w^*|w^*_n\geq w_n(q_n)\}$ is
approachable. Next we prove that $[w_n(q_n)]$ is the only feasible
vector in the set. Consider any vector $w'\neq w(q)$ in the set. We
have $w'_n\geq w_n(q_n)$ for all $n$, and $w'_{n_0}>
w_{n_0}(q_{n_0})$, for some $n_0$. Suppose $n_0\in H_{k}\backslash
H_{k-1}$. We have:
\begin{align*}
\sum_{n\in H_k}w'_n>&\sum_{n\in H_k}w_n(q_n)=\sum_{i=1}^k\sum_{n\in
H_i\backslash
H_{i-1}}\tau\frac{b_n+\theta_k}{a_n}\\
=&\mbox{$\sum_{i=1}^k$} (I_{H_{i-1}}-I_{H_{i}})=\tau - I_{H_k},
\end{align*}
and thus $w'$ is not feasible. Therefore, $w(q)$ is the only
feasible vector in $\{w^*|w^*_n\geq w_n(q_n)\}$, and the vector of
work performed resulting from the generalized transmission time
policy must converge to $w(q)$.
\end{IEEEproof}

\begin{corollary}
For the policy of Theorem \ref{theorem:convergence}, the vector of
delivery ratios converges to $q$.
\end{corollary}
\begin{IEEEproof}
Follows from Lemma \ref{lemma:workload}.
\end{IEEEproof}

\subsection{Optimality of the Weighted Transmission Policy for $ACCESS$-$POINT$}   \label{subsection:optimal}

\begin{theorem} \label{theorem:scheduling}
Given $[\rho_n]$, the vector $q$ of long-term average delivery
ratios resulting from the WT policy is a solution to
$ACCESS$-$POINT$.
\end{theorem}
\begin{IEEEproof}
We use the sequence of sets $\{H_k\}$ and values $\{\theta_k\}$,
with $a_n:=\frac{1}{\rho_n}$ and $b_n:=0$, as defined in the proof
of Theorem \ref{theorem:convergence}. Let $K:=|\{\theta_k\}|$. Thus,
we have $H_K=TOT=\{1,2,\dots, N\}$. Also, let $m_k:= |H_k|$. For
convenience, we renumber clients so that $H_k=\{1,2,\dots,m_k\}$.
The proof of Theorem \ref{theorem:convergence} shows that $q_n=\tau
p_n\theta_k\rho_n$, for $n\in H_k\backslash H_{k-1}$. Therefore,
$w_n(q_n)=\frac{q_n}{p_n}=\tau\theta_k\rho_n$. Obviously, $q$ is
feasible, since it is indeed achieved by the WT policy. Thus, to
establish optimality, we only need to prove the existence of vectors
$\zeta$ and $\mu$ that satisfy conditions
(\ref{NETWORK:KKT1})--(\ref{NETWORK:KKT4}).

Set $\mu_n=0$, for all $n$. Let
$\zeta_{H_K}=\zeta_{TOT}:=\frac{\rho_N}{w_N(q_N)}=\frac{1}{\tau\theta_K}$
and
$\zeta_{H_k}:=\frac{\rho_{m_{k}}}{w_{m_{k}}(q_{m_{k}})}-\frac{\rho_{m_{k+1}}}{w_{m_{k+1}}(q_{m_{k+1}})}=\frac{1}{\tau\theta_k}-\frac{1}{\tau\theta_{k+1}}$,
for $1\leq k \leq K-1$. Finally, let $\zeta_S:=0$, for all
$S\notin\{H_1,H_2,\dots, H_K\}$. We claim that the vectors $\zeta$
and $\mu$, along with $q$, satisfy conditions
(\ref{NETWORK:KKT1})--(\ref{NETWORK:KKT4}).

We first evaluate condition (\ref{NETWORK:KKT1}). Suppose client $n$
is in $H_k\backslash H_{k-1}$. Then client $n$ is also in $H_{k+1},
H_{k+2},\dots,H_K$. So,
\begin{align*}
&-\frac{\rho_n}{q_n}+\frac{\sum_{S\ni n} \zeta_S}{p_n}-\mu_n
=-\frac{1}{\tau\theta_k p_n}+\frac{\sum_{i=k}^K
\zeta_{H_i} }{p_n}\\
=&-\frac{1}{\tau\theta_k p_n}+\frac{1}{\tau\theta_k p_n}=0.
\end{align*}
Thus, condition (\ref{NETWORK:KKT1}) is satisfied.

Since $\mu_n=0$, for all $n$, condition (\ref{NETWORK:KKT3}) is
satisfied. Further, since
$\frac{1}{\theta_k}>\frac{1}{\theta_{k+1}}$, for all $1\leq k\leq
K-1$, condition (\ref{NETWORK:KKT4}) is also satisfied. It remains
to establish condition (\ref{NETWORK:KKT2}). Since $\zeta_S=0$ for
all $S\notin\{H_1, H_2,\dots, H_K\}$, we only need to show
$\sum_{i\in S} \frac{q_i}{p_i}-(\tau-I_{S})=0$ for
$S\in\{H_1,H_2,\dots, H_K\}$.

Consider $H_k$. For each client $i\in H_k$ and each client $j\notin
H_k$, $\frac{w_i(q_i)}{\rho_i} < \frac{w_j(q_j)}{\rho_j}$. Since
$w_n(q_n)$ is the average number of time slots that the AP spends on
working for client $n$, we have $\frac{u_i(t)}{\rho_i} <
\frac{u_j(t)}{\rho_j}$, for all $i\in H_k$ and $j\notin H_k$, after
a finite number of periods. Therefore, except for a finite number of
periods, clients in $H_k$ will have priorities over those not in
$H_k$. In other words, if we only consider the behavior of those
clients in $H_k$, it is the same as if the AP only works on the
subset $H_k$ of clients. Further, recall that $I_{H_k}$ is the
expected number of time slots that the AP is forced to stay idle
when the AP only works on the subset $H_k$ of clients. Thus, we have
$\sum_{i\in H_k}w_i(q_i)=\tau-I_{H_k}$ and $\sum_{i\in H_k}
\frac{q_i}{p_i}-(\tau-I_{H_k})=0$, for all $k$.
\end{IEEEproof}

\subsection{Fairness of Allocated Delivery Ratios} \label{subsection:fairness}
We now show that the WT policy not only solves the $ACCESS$-$POINT$
problem but also achieves some forms of fairness among clients. Two
common fairness criteria are \emph{max-min fair} and
\emph{proportionally fair}. We extend the definitions of these two
criteria as follows:

\begin{definition}  \label{definition:max-min fair}
A scheduling policy is called \emph{weighted max-min fair with
positive weight vector $a=[a_n]$} if it achieves $q$, and, for any
other feasible vector $q'$, we have: $q'_i > q_i \Rightarrow q'_j <
q_j,$ for some $j$ such that $\frac{w_i(q_i)}{a_i} \geq
\frac{w_j(q_j)}{a_j}$.
\end{definition}

\begin{definition}  \label{definition:proportional fair}
A scheduling policy is called \emph{weighted proportionally fair
with positive weight vector $a$} if it achieves $q$ and, for any
other feasible vector $q'$, we have:
\[
\mbox{$\sum_{n=1}^N$} \frac{w_n(q'_n) - w_n(q_n)}{w_n(q_n)/a_n}\leq
0.
\]
\end{definition}

Next, we prove that the WT policy is both weighted max-min fair and
proportionally fair with weight vector $\rho$.

\begin{theorem} \label{theorem:max-min fair}
The weighted transmission policy is weighted max-min fair with
weight $\rho$
\end{theorem}
\begin{IEEEproof}
We sort clients and define $\{H_k\}$ as in the proof of Theorem
\ref{theorem:scheduling}. Let $q$ be the vector achieved by the WT
policy and $q'$ be any feasible vector. Suppose $q'_i > q_i$ for
some $i$. Assume client $i$ is in $H_k\backslash H_{k-1}$. The proof
in Theorem \ref{theorem:scheduling} shows that $\sum_{n\in H_k}
w_n(q_n)=\tau - I_{H_k}$. On the other hand, the feasibility
condition requires $\sum_{n\in H_k} w_n(q'_n) \leq \tau -
I_{H_k}=\sum_{n\in H_k} w_n(q_n)$. Further, since $q'_i > q_i$,
$w_i(q'_i) > w_i(q_i)$, there must exist some $j\in H_k$ so that
$w_j(q'_j) < w_j(q_j)$, that is, $q'_j < q_j$. Finally, since $i\in
H_k\backslash H_{k-1}$, we have $\frac{w_i(q_i)}{\rho_i}\geq
\frac{w_n(q_n)}{\rho_n}$, for all $n\in H_k$, and hence
$\frac{w_i(q_i)}{\rho_i}\geq \frac{w_j(q_j)}{\rho_j}$.
\end{IEEEproof}

\begin{theorem} \label{theorem:proportional fair}
The weighted transmission policy is proportionally fair with weight
$\rho$.
\end{theorem}
\begin{IEEEproof}
We sort clients and define $\{H_k\}$ as in the proof of Theorem
\ref{theorem:scheduling}. Let $q$ be the vector achieved by the WT
policy, and let $q'$ be any feasible vector. We have
$\frac{w_i(q_i)}{\rho_i}=\tau\theta_k$, if $i\in H_k\backslash
H_{k-1}$. Define $\Delta_k:=\sum_{n\in H_k\backslash
H_{k-1}}w_n(q'_n) - w_n(q_n)$.

To prove the theorem, we prove a stronger statement by induction:
\[
\sum_{n\in H_k}\frac{w_n(q'_n) -
w_n(q_n)}{w_n(q_n)/\rho_n}=\sum_{i=1}^k\frac{\Delta_i}{\tau\theta_i}\leq
0, \mbox{ for all }k>0.
\]

First consider the case $k=1$. The proof in Theorem
\ref{theorem:scheduling} shows that $\sum_{n\in H_1} w_n(q_n)=\tau -
I_{H_1}$. Further, the feasibility condition requires $\sum_{n\in
H_1} w_n(q'_n)\leq\tau - I_{H_1}=\sum_{n\in H_1} w_n(q_n)=\tau -
I_{H_1}$, and so $\Delta_1=\sum_{n\in H_1} w_n(q'_n)-w_n(q_n)\leq
0$. Thus, we have $\frac{\Delta_1}{\tau\theta_1}\leq 0$.

Suppose we have $\sum_{i=1}^{k} \frac{\Delta_i}{\tau\theta_i}\leq
0,$ for all $k\leq k_0$. Again, the proof in Theorem
\ref{theorem:scheduling} gives us $\sum_{n\in H_{k_0+1}}
w_n(q_n)=\tau - I_{H_{k_0+1}}$ and the feasibility condition
requires $\sum_{n\in H_{k_0+1}} w_n(q'_n)\leq\tau -
I_{H_{k_0+1}}=\sum_{n\in H_{k_0+1}} w_n(q_n)$. Thus,
$\sum_{i=1}^{k_0+1} \Delta_i \leq 0.$ We can further derive:
\begin{align*}
&\sum_{i=1}^{k_0+1}\frac{\Delta_i}{\tau\theta_i}\\
\leq
&\sum_{i=1}^{k_0}
\frac{\Delta_i}{\tau\theta_i}(1-\frac{\theta_i}{\theta_{k_0+1}})\hskip
15pt (\mbox{since }
\sum_{i=1}^{k_0+1}\frac{\Delta_i}{\tau\theta_{k_0+1}}\leq 0)\\
=&\sum_{j=1}^{k_0}[(\frac{\theta_{j+1}-\theta_j}{\theta_{k_0+1}})\sum_{i=1}^j
\frac{\Delta_i}{\tau\theta_i}]\\
\leq&0\hskip 15pt (\mbox{since }
\sum_{i=1}^j\frac{\Delta_i}{\tau\theta_i}\leq 0,\mbox{ and }
\theta_{j+1}>\theta_j, \forall j\leq k_0)
\end{align*}

By induction, $\sum_{i=1}^k\frac{\Delta_i}{\tau\theta_i}\leq 0$, for
all $k$. Finally, we have:
\[
\sum_{n=1}^N \frac{w_n(q'_n) -
w_n(q_n)}{w_n(q_n)/\rho_n}=\sum_{i=1}^K
\frac{\Delta_i}{\tau\theta_i}\leq 0,
\]
and the WT policy is proportionally fair with weight $\rho$.
\end{IEEEproof}

\section{Simulation Results}
\label{section:simulation}

We have implemented the WT policy and the bidding game, as described
in Section \ref{section:bidding}, on ns-2. We use the G.711 codec
for audio compression to set the simulation parameters, as
summarized in Table \ref{table:parameter}. All results in this
section are averages of 20 simulation runs.

\begin{table}[h]
  \centering
  \caption{Simulation Setup}\label{table:parameter}
  \begin{tabular}{|l|l|}
    \hline
    Packetization interval & 20 $ms$ \\
    \hline
    Payload size per packet & 160 Bytes \\
    \hline
    Transmission data rate & 11 Mb/s\\
    \hline
    Transmission time (including MAC overheads) & 610 $\mu s$\\
    \hline
    \# of time slots in a period & 32\\
    \hline
  \end{tabular}
\end{table}

\subsection{Convergence Time for the Weighted Transmission Policy}
We have proved that the vector of delivery ratios will converge
under the WT policy in Section \ref{subsection:convergence}.
However, the speed of convergence is not discussed. In the bidding
game, we assume that the delivery ratio observed by each client is
post convergence. Thus, it is important to verify whether the WT
policy converges quickly. In this simulation, we assume that there
are 30 clients in the system. The $n^{th}$ client has channel
reliability $(50+n)\%$ and offers a bid $\rho_n=(n\mod 2)+1$. We run
each simulation for 10 seconds simulation time and then compare the
absolute difference of $\sum_n \rho_n\log q_n$ between the delivery
ratios at the end of each period with those after 10 seconds. In
particular, we artificially set $q_n=0.001$ if the delivery ratio
for client $n$ is zero, to avoid computation error for $\log q_n$.

\begin{figure}
\includegraphics[width=3.2in]{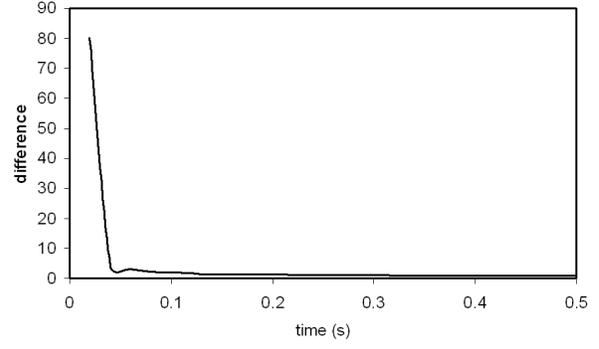}
\caption{Convergence of the weighted transmission
policy}\label{fig:converge}
\end{figure}

 Simulation results are shown in Fig.
\ref{fig:converge}. It can be seen that the delivery ratios converge
rather quickly. At time 0.2 seconds, the difference is smaller than
$1.4$, which is less than 10$\%$ of the final value. Based on this
observation, we assume that each client updates its bid every 0.2
seconds in the following simulations.

\subsection{Utility Maximization}
In this section, we study the total utility that is achieved by
iterating between the bidding game and the WT policy, which we call
WT-Bid. We assume that the utility function of each client $n$ is
given by $\gamma_n\frac{q_n^{\alpha_n}-1}{\alpha_n}$, where
$\gamma_n$ is a positive integer and $0<\alpha_n<1$. This utility
function is strictly increasing, strictly concave, and
differentiable for any $\gamma_n$ and $\alpha_n$. In addition to
evaluating the policy WT-Bid, we also compare the results of three
other policies: a policy that employs the WT policy but without
updating the bids from clients, which we call WT-NoBid; a policy
that decides priorities randomly among clients at the beginning of
each period, which we call Rand; and a policy that gives clients
with larger $\gamma_n$ higher priorities, with ties broken randomly,
which we call P-Rand.

In each simulation, we assume there are 30 clients. The $n^{th}$
client has channel reliability $p_n=(50+n)\%$, $\gamma_n=(n\mod
3)+1$, and $\alpha_n=0.3+0.1(n\mod 5)$. In addition to plotting the
average of total utility over all simulation runs, we also plot the
variance of total utility.

\begin{figure}
\subfloat[Average of total utility]{
\label{fig:average} %% label for first subfigure
\includegraphics[width=3.2in]{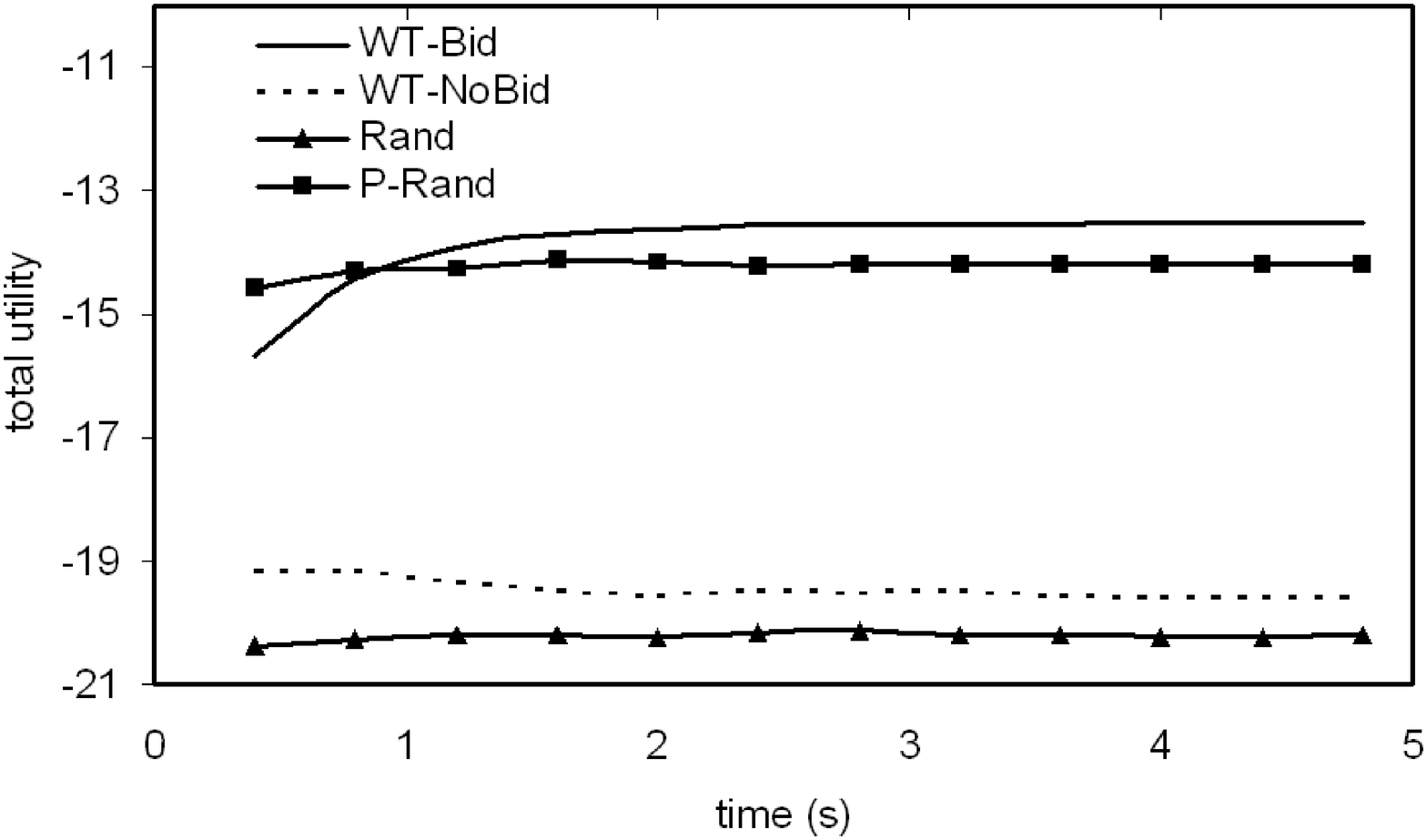}}
\hspace{0.01\linewidth} \subfloat[Variance of total utility]{
\label{fig:variance} %% label for second subfigure
\includegraphics[width=3.2in]{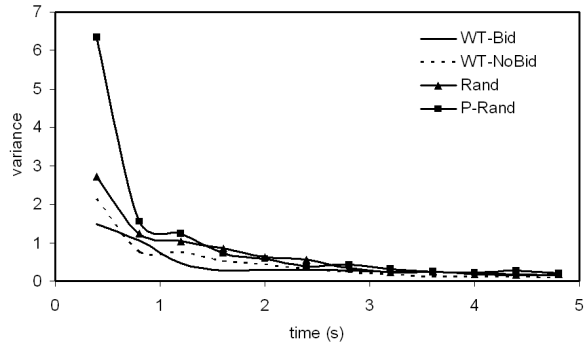}}
\caption{Performance of total utility}\label{fig:utility}
\end{figure}

Fig. \ref{fig:utility} shows the simulation results. The WT-Bid
policy not only achieves the highest average total utility but also
the smallest variance. This result suggests that the WT-Bid policy
converges very fast. On the other hand, the WT-NoBid policy fails to
provide satisfactory performance since it does not consider the
different utility functions that clients may have. The P-Rand policy
offers much better performance than both the WT-NoBid policy and the
Rand policy since it correctly gives higher priority to clients with
higher $\gamma_n$. Still, it cannot differentiate between clients
with the same $\gamma_n$ and thus can only provide suboptimal
performance.

\section{Concluding Remarks}    \label{section:conclusion}

We have studied the problem of utility maximization problem for
clients that demand delay-based QoS support from an access point.
Based on an analytical model for QoS support proposed in previous
work, we formulate the utility maximization problem as a convex
optimization problem. We decompose the problem into two simpler
subproblems, namely, $CLIENT_n$ and $ACCESS$-$POINT$. We have proved
that the total utility of the system can be maximized by jointly
solving the two subproblems. We also describe a bidding game to
reconciliate the two subproblems. In the game, each client announces
its bid to maximize its own net profit and the AP allocates time
slots to achieve the optimal point of $ACCESS$-$POINT$. We have
proved that the equilibrium point of the bidding game jointly solves
the two subproblems, and therefore achieves the maximum total
utility.

In addition, we have proposed a very simple, priority-based weighted
transmission policy for solving the $ACCESS$-$POINT$ subproblem.
This policy does not require that the AP know the channel
reliabilities of the clients, or their individual utilities. We have
proved that the long-term performance of a general class of
priority-based policies that includes our proposed policy converges
to a single point. We then proved that the limiting point of the
proposed scheduling policy is the optimal solution to
$ACCESS$-$POINT$. Moreover, we have also proved that the resulting
allocation by the AP satisfies some forms of fairness criteria.
Finally, we have implemented both the bidding game and the
scheduling policy in ns-2. Simulation results suggests that the
scheduling policy quickly results in convergence. Further, by
iterating between the bidding game and the WT policy, the resulting
total utility is higher than other tested policies.

\def\baselinestretch{0.93}
\small
\bibliographystyle{plain}
\bibliography{reference}

\end{document}